# Rapid Polymorphic Screening using Sessile Microdroplets: Competitive Nucleation of Mannitol Polymorphs

Ruel Cedeno[1], Romain Grossier[1], Nadine Candoni[1], Stéphane Veesler[1]*

[1]CNRS, Aix-Marseille Université, CINaM (Centre Interdisciplinaire de Nanosciences de Marseille), Campus de Luminy, Case 913, F-13288 Marseille Cedex 09, France

**Abstract**

We developed a rapid polymorphic screening approach based on contracting sessile microdroplets which offers several advantages: (1) achieves very high supersaturation to facilitate formation of metastable forms (2) allows systematic labeling of samples (3) gives access to statistical distribution of polymorphic selectivity as a function of experimental conditions (4) ensures the formation of crystal for each droplet, addressing the problem of uncrystallized droplets in traditional microfluidics.

We studied the competitive nucleation of D-mannitol polymorphs and investigated the effect of droplet volume on polymorphic selectivity. We showed that our observed polymorph distributions at different volumes are qualitatively consistent with the predictions of classical nucleation theory except for very small volumes where thermodynamic confinement or surface effects could play a substantial role.

Overall, our microfluidic approach can be a promising tool not only for routine screening of pharmaceutical polymorphs in the industrial context but also in the fundamental understanding of the mechanisms underlying the competitive nucleation of polymorphs.

**Introduction**

Polymorphism, the ability of a material to form multiple crystal structures, plays a crucial role in numerous fields, notably in the pharmaceutical industry in which the solid forms directly impact the bioavailability, processability, and overall quality of the final product.[1] For this reason, understanding the kinetics of competitive polymorphic nucleation is essential in controlling the polymorphic outcome.[2] Given that nucleation is inherently stochastic[3], experimental platforms allowing large number of experiments are needed to access the probability distribution of polymorphic selectivity as a function of various experimental conditions.[4-6]

In our previous work, we have developed a microfluidic platform that allows extraction of thermodynamic and kinetic parameters of nucleation in contracting sessile microdroplets.[3, 7] In this work, we further extend its application in the study of polymorphism. Using D-mannitol in water as a model system, we demonstrate how our platform enables rapid screening of polymorphic outcomes. Moreover, it also allows statistical investigation of the influence of volume and supersaturation ratio $S$ (i.e. $c_{sol}/c_{eq}$ where $c_{sol}$ is the solution concentration and $c_{eq}$ is the saturation concentration of the β polymorph) on polymorphic selectivity. We then rationalize our observed polymorphic distributions using classical nucleation theory, revealing interesting insights into the interplay of thermodynamics and kinetics in the stochastic nucleation of polymorphs.

**Materials and methods**

The D-mannitol (≥ 98%), a crystalline excipient, was purchased from Sigma Aldrich and tested as the β form by Raman spectroscopy. The β form is the thermodynamically stable polymorph of mannitol, and the α and δ forms are metastable (stability order: $\delta < \alpha < \beta$ at room temperature)[8]. CCDC numbers are 224658, 224659 and 22460 for polymorphs α, β and $\delta$ respectively.

Microdroplet generation and nucleation time detection technique are based on a previously reported experimental setup and protocol in Ref[7], consisting in deliquescence/recrystallization cycling.

However, unlike NaCl that dissolves upon exposure to humidity (RH > 75%), D-mannitol does not absorb enough moisture to undergo deliquescence, preventing the use of the RH cycling technique. We thus modified the procedure by generating initially undersaturated arrays of sessile microdroplets on PMMA coated glass immersed in a thin layer of polydimethylsiloxane (PDMS) oil (10 cSt) at ambient conditions (1 atm, 25°C). This is done at RH close to 100% to minimize possible evaporation during microdroplet generation. Once the desired number of microdroplets are generated, the RH is lowered to 10%, causing the droplets to evaporate by diffusion of water in oil and eventually nucleate. Using a tailor-made evaporation model[9] (details in section S1 in the ESI), the supersaturation ratio of the microdroplets can be obtained as a function of time. The crystallized droplets are analysed in situ at the end of the experiment using a Kaiser RXN1 Raman microscope system. Measurements are made at room temperature using a 785-nm laser.

## Results and discussion

**Polymorph Screening and Characterization.** To demonstrate how our sessile microdroplet approach can accelerate polymorph screening, we used D-mannitol in water as a case study. The bottom-view image of generated sessile microdroplets is shown in **Figure 1a** (see the video in the ESI for the whole process, droplet contraction until crystallization). Supersaturation at generation is $S_0$ = 0.25.

The resulting crystal from each microdroplet were then characterized using Raman spectroscopy. Three distinct spectra were obtained corresponding to the α,β,δ forms as shown in **Figure 1b.** These spectra exhibit sufficiently resolved characteristic peaks for precise phase identification.

In contrast to conventional screening methods, our approach offers several advantages: (1) The relatively high supersaturation level achieved in microdroplets facilitates the formation of metastable polymorphs (2) The linear arrangement of the immobilized crystals facilitates systematic labeling of each sample with respect to its position in the 2D array, allowing us to generate a cartography of polymorph shown in **Figure 1c**. (3) The statistical distribution of polymorphic selectivity can be analyzed as a function of supersaturation ratio $S$ as shown in **Figure 1d**. We use image acquisition to track the droplets as they evaporate, so we know when nucleation occurred and at what concentration. We use Raman spectroscopy to characterize the nucleated phase. (4) The open geometry microfluidics ensures the formation of crystal for each droplet, unlike closed microfluidics (chips or tubing) where droplets can remain metastable for several days to weeks. (5) The method is relatively rapid, as the entire experiment from droplet generation to characterization took less than 4 hours in this work.

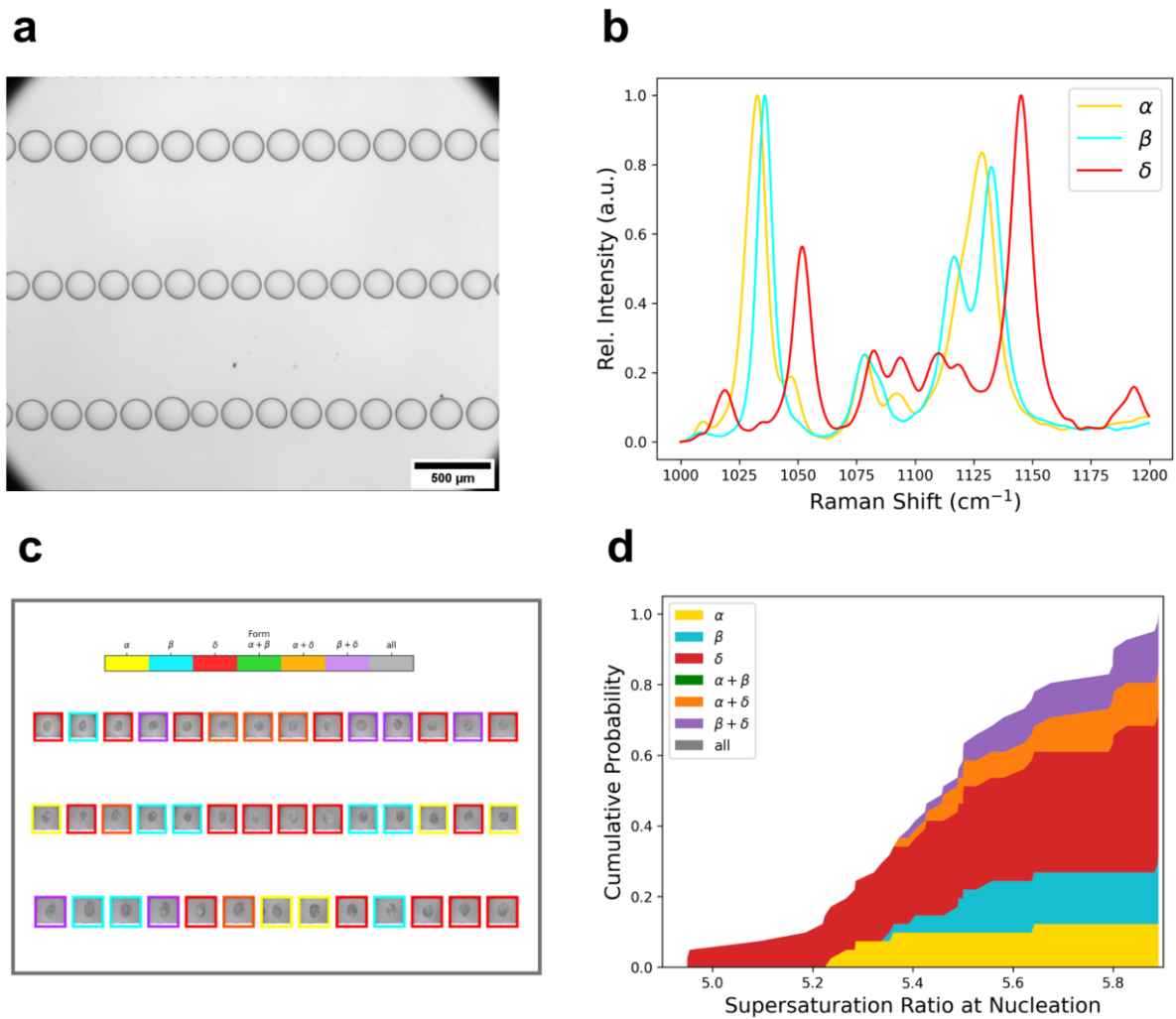

**Figure 1. (a)** Bottom-view image of aqueous D-mannitol microdroplet arrays at generation ($S_0 = 0.25$), the volume at saturation is 0.7nL **(b)** Raman spectra corresponding to α,β,δ forms **(c)** cartography of polymorphic outcome with its corresponding crystal image **(d)** polymorphic distribution as a function of supersaturation ratio at nucleation.

**Influence of Droplet Volume.** In this communication, we focus on the influence of droplet volume on cartography of polymorphism. The droplet volume is an important parameter in microfluidic crystallization as it has a direct impact on the surface area to volume ratio as well as the achievable supersaturation level (which influence the relative importance of homogeneous and heterogeneous nucleation mechanisms), and on the degree of thermodynamic confinement (i.e. reduction of effective supersaturation upon formation of pre-critical clusters in small volumes)[10]. Thus, it would be interesting to understand how the droplet volume impacts the polymorphic selectivity of D-mannitol.

The polymorphic distribution for various volumes at saturation (0.2, 0.3, 0.7, 1.5, and 5 nL) is shown in **Figure 2a**. Notice that the physical mixtures (α+β, α+δ, β+δ) occupy a not negligible fraction across different volumes. The formation of polymorphic mixture can be interpreted in three different ways:
Case 1: Independent nucleation: mixtures arise from the independent/concomitant nucleation event of each polymorph.
Case 2: Ostwald rule of stages: The less stable polymorph nucleates first then gradually transforms to a more stable polymorph.

Case 3: Surface nucleation: The more stable polymorph nucleates first, which then facilitates the nucleation of the less stable ones on its surface.

It should be noted that the method used to measure induction times makes it possible to observe simultaneously hundreds of droplets at the cost of a loss of resolution[11] , which makes it impossible to distinguish between the 3 cases described above.

The resulting distribution based on Case 1, Case 2, and Case 3 representations are plotted in Figure 2b, 2c, and 2d respectively, considering only the first polymorph appeared according to case 2 or 3 respectively.

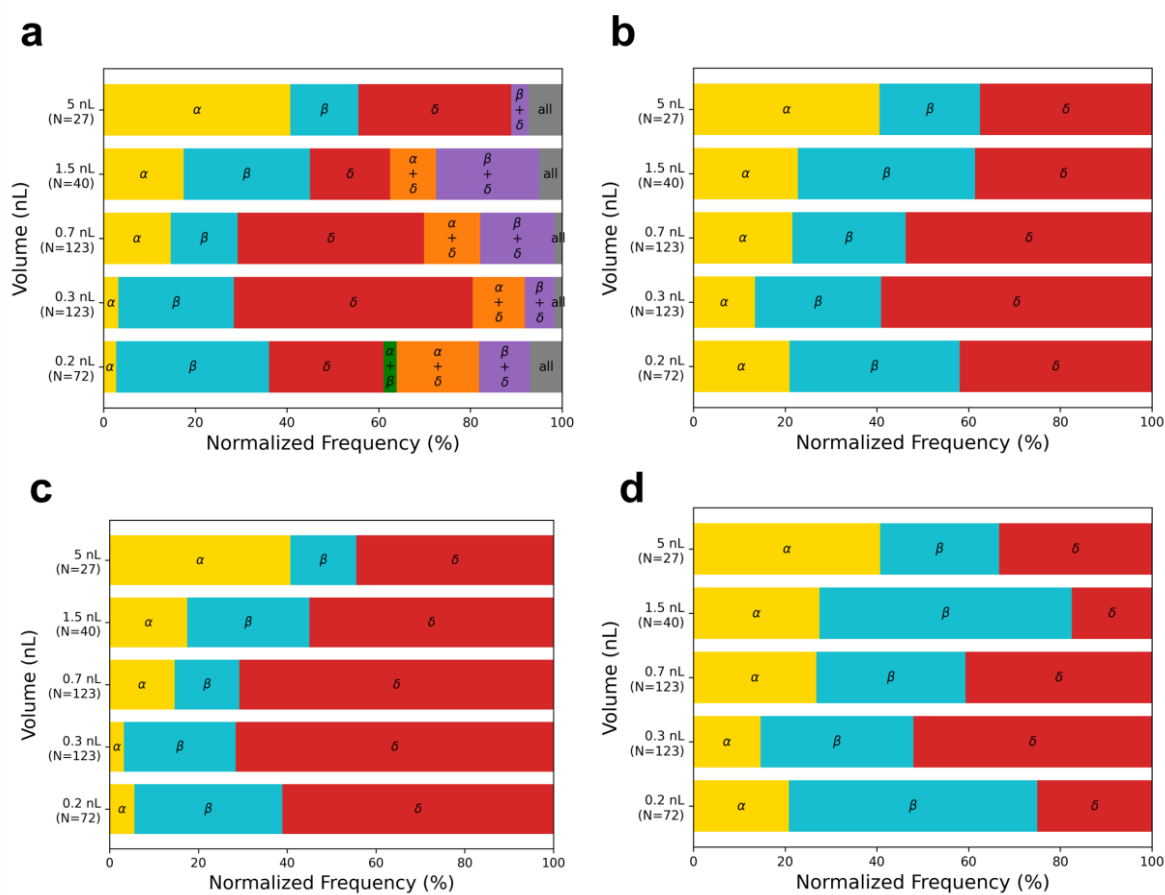

**Figure 2**. Distribution of polymorphs as a function of droplet volume in which mixtures are represented as (a) actual mixtures (b) Case 1 (c) Case 2 (d) Case 3.  N indicates the number of droplets.

Among these three mixture interpretations, we believe Case 3 is highly unlikely because as the stable form nucleates and grows, it quickly depletes the supersaturation level in the droplet which consequently dissolves the precursors/pre-critical clusters of the more soluble metastable forms. Therefore, in the following discussion, we will concentrate on cases 1 and 2, for which it is impossible to give a definitive answer.

Recall that the order of stability for D-mannitol polymorphs is δ < α < β. During droplet contraction, the experimental conditions inside each droplet move on the phase diagram from undersaturated to supersaturated. They first cross the solubility curve of form β, before crossing the solubility curve of forms α and then δ. The same trend is expected for the metastable limit of the three forms. Therefore, we would expect that the stable β form will dominate at larger volumes. This is due to the lower surface area to volume ratio of larger droplets, which implies that their supersaturation ratio S increases more slowly (fig.S2b in the ESI), allowing more time in the stable β-form nucleation area. While the least stable δ form will dominate at smaller volumes due to the higher surface area to volume ratio of smaller droplets, and so their supersaturation ratio S increases more rapidly (fig.S2b in the ESI). Hence, metastable products are favoured according to Ostwald's rule of stages, as previously observed by Myerson group in confined environment for sulfathiazole and glycine[12, 13] and by Buanz et al. for crystallization in printed droplets of D-mannitol.[14] As shown in **Figure 2b-2c**, the trend in % δ-form qualitatively agrees with our hypothesis, i.e., it tends to increase as the volume decreases (except at 0.2 nL). Surprisingly, the % β-form does not increase with volume, and the least stable δ-form dominates across all studied volumes. Interestingly, the medium stable α-polymorph follows the expected trend of the stable polymorph.

**Predictions of Classical Nucleation Theory.** To gain insights into the observed dominance of δ-form, we used the classical nucleation theory (CNT) to model the polymorphic selectivity as a function of supersaturation. Recall that CNT requires two parameters, pre-exponential factor A (kinetic parameter) and the effective interfacial energy (between crystal and solution) $\gamma_{eff}$ (thermodynamic parameter). In linearized form of the equation for heterogeneous nucleation, CNT can be written as:

$$\ln\left(\frac{J}{A}\right) = -\frac{16\pi}{3} \frac{\gamma_{eff}^3}{\rho_s^2 (k_b T)^3 \ln^2 S} \qquad (1)$$

where $J$ is the nucleation rate, $\rho_s$ is the number density (formula units per m³) and $k_b T$ is the thermal energy.

For simplicity, given that the mass transfer properties (i.e. diffusivity, viscosity, etc) in the liquid phase are identical regardless of the solid form, we can suppose that the parameter *A* (related to the mass transfer rate towards the nucleus) is similar for all polymorphs. This assumption is similar to that of Sato[15] and Deij et al.[16] Consequently, the relative nucleation rate ln(J/A) is only a function of $\gamma_{eff}$. In principle, $\gamma_{eff}$ can be measured from the probability distributions of nucleation time for each polymorph. Unfortunately, given the dominance of δ-form, the number of data points for other polymorphs is not sufficient to extract reliable statistical distribution. For this reason, we decided to apply the empirical correlation of Mersmann[17] which correlates the interfacial energy (between crystal and solution) $\gamma_{SL}$ with the solubility, written as

$$\gamma_{SL} = K(c_{sol})^{2/3} \ln\left(\frac{c_{sol}}{c_{eq}}\right) \text{ where } K = 0.414 k_b T \, N_A^{2/3} \qquad (2)$$

with *K* as an empirical parameter, and $c_{sol}$ and $c_{eq}$ as the equilibrium solid-state and liquid-phase concentrations (in mol/m³) respectively and $N_A$ is the Avogadro's number. $\gamma_{SL}$ is used for homogeneous nucleation in the CNT equation (1), for more details see ref[18]. Using the solubility data of Su et al[19] (values in table S1 in the ESI) together with equation (2), we obtained $\gamma_{SL}$ as 9.6, 8.6 and 8.1 mJ/m² for β, α and δ respectively, in agreement with the rule the smaller the interfacial energy the larger the solubility. The resulting relative nucleation rate ln(*J/A*) and the relative polymorphic selectivity (i.e. the probability of forming the polymorph, equation 5 of Ref[20]) is plotted in **Figure 3a** and **3b** respectively.

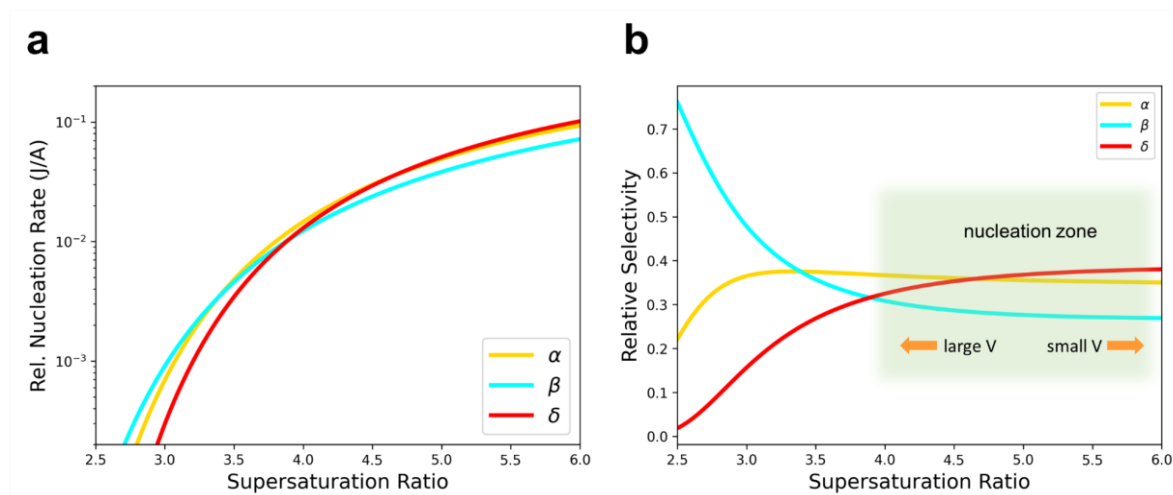

**Figure 3**. **(a)** Relative nucleation rate and **(b)** relative selectivity as a function of supersaturation ratio computed based on classical nucleation theory and Mersmann correlation. The green area in the graph corresponds to the estimated supersaturation at nucleation in the different experiments presented in Figure 2.

To interpret the results in **Figure 3**, remember that nucleation time is inversely proportional to the droplet volume ($t_n=1/JV$). Under continuous evaporation, smaller droplets can therefore achieve higher supersaturations within the nucleation zone. Thus, **Figure 3b** suggests that larger droplets would favor the α-form while the smaller ones would favor the δ-form. Reassuringly, this behavior is coherent with what we observe in Figure 2b and 2c, the predominance of polymorphs α and δ, except that of the 0.2 nL dataset. While one could speculate several possible explanations, such change in trend at very low volumes could be likely due to confinement effects[10] which lowers the effective supersaturation and consequently promotes the α-form. Moreover, diminishing droplet volume provides higher ratio surface to volume, potentially affecting nucleation mechanisms and rates.

Indeed, our microfluidic platform and modeling approach reveal interesting insights into the impact of volume on the stochastic nucleation of mannitol polymorphs. Moreover, our approach can be extended to study other polymorphic material of interest.

**Conclusions**

In this work, we developed a rapid polymorphic screening approach based on contracting sessile microdroplets which offers several advantages: (1) achieves very high supersaturation to facilitate formation of metastable forms (2) allows systematic labeling of samples (3) gives access to statistical distribution of polymorphic selectivity as a function of experimental conditions (4) ensures the formation of crystal for each droplet, addressing the problem of uncrystallized droplets in traditional microfluidics.

Thanks to this platform, we studied the competitive nucleation of D-mannitol polymorphs and investigated the effect of droplet volume on polymorphic selectivity. We showed that our observed polymorph distributions at different volumes are qualitatively consistent with the predictions of the classical nucleation theory except for very small volumes where thermodynamic confinement or surface effects could play a substantial role.

Overall, our microfluidic approach can be a promising tool not only for routine screening of pharmaceutical polymorphs in the industrial context but also in the fundamental understanding of the mechanisms underlying the competitive nucleation of polymorphs.

## Author contributions

**Ruel Cedeno** : Formal analysis (equal); Methodology (equal); Writing – review & editing (equal). **Romain Grossier** : Supervision (equal); Formal analysis (equal); Conceptualization (equal); Writing – review & editing (equal). **Nadine Candoni** : Supervision (equal); Conceptualization (equal); Writing – original draft (equal). **Stéphane Veesler** : Supervision (equal); Conceptualization (equal); Writing – original draft (equal).

## Data availability

Data for this article are available at zenodo at https://doi.org/10.5281/zenodo.12665873 .

## Acknowledgements

RC acknowledges the financial support of ANR-FACET, ANR-19-CE08-0014-02.

Supplementary Material for:

# Rapid Polymorphic Screening using Sessile Microdroplets: Competitive Nucleation of Mannitol Polymorphs

Ruel Cedeno[1], Romain Grossier[1], Nadine Candoni[1], Stéphane Veesler[1]*

[1]CNRS, Aix-Marseille Université, CINaM (Centre Interdisciplinaire de Nanosciences de Marseille), Campus de Luminy, Case 913, F-13288 Marseille Cedex 09, France

**S1. Evaporation Model**

**Table S1** Numerical values used as input in evaporation model[1] for aqueous D-mannitol droplets and Mersmann correlation[2], referenced at 25°C and 1 atm.

| Quantity | Symbol | Value | Unit |
|---|---|---|---|
| solubility of water in PDMS oil[1] | $c_s$ | 8.76 | mol/m$^3$ |
| diffusivity of water in PDMS oil[1] | $D$ | 6.74 × 10$^{-9}$ | m$^2$s$^{-1}$ |
| supersaturation at matching time* | $S_{match}$ | 4.00 | - |
| coefficient of density change for mannitol[3] | $b_1$ | 0.0719 | - |
| coefficient of water activity lowering for mannitol[4] | $b_2$ | 0.194 | - |
| solubility of β-mannitol in water[5] | $c_{eq}$ | 1.185 | mol/kg |
| solubility of β-mannitol in water[5] | $c_{eq}$ | 1217.4 | mol/m$^3$ |
| solubility of α-mannitol in water[5] | $c_{eq}$ | 1453.0 | mol/m$^3$ |
| solubility of δ-mannitol in water[5] | $c_{eq}$ | 1667.5 | mol/m$^3$ |
| equilibrium solid-state β-mannitol[5] | $C_{sol}$ | 8323.7 | mol/m$^3$ |
| equilibrium solid-state α-mannitol[5] | $C_{sol}$ | 8210.5 | mol/m$^3$ |
| equilibrium solid-state δ-mannitol[5] | $C_{sol}$ | 8365.6 | mol/m$^3$ |
| molar mass of mannitol | $M_{Man}$ | 0.182 | kg/mol |
| diffusivity of mannitol in water[6] | $D_i$ | 6.05×10$^{-10}$ | m$^2$/s |
| density of pure water[7] | $\rho_w$ | 997 | kg/m$^3$ |

*measured by monitoring/interpolating the droplet volume (lateral view) as it optically disappears, supersaturation $S = c/c_{eq}$

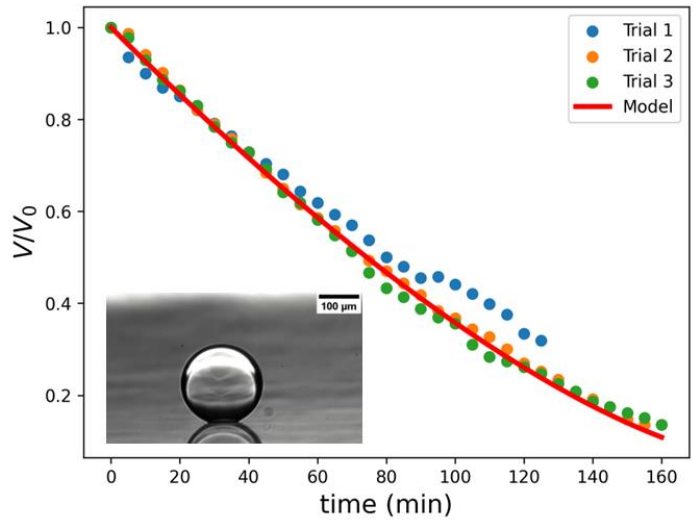

**Figure S1.** Evolution of relative droplet volume ($V/V_0$) as a function of time (3 replicates) taken from lateral images.

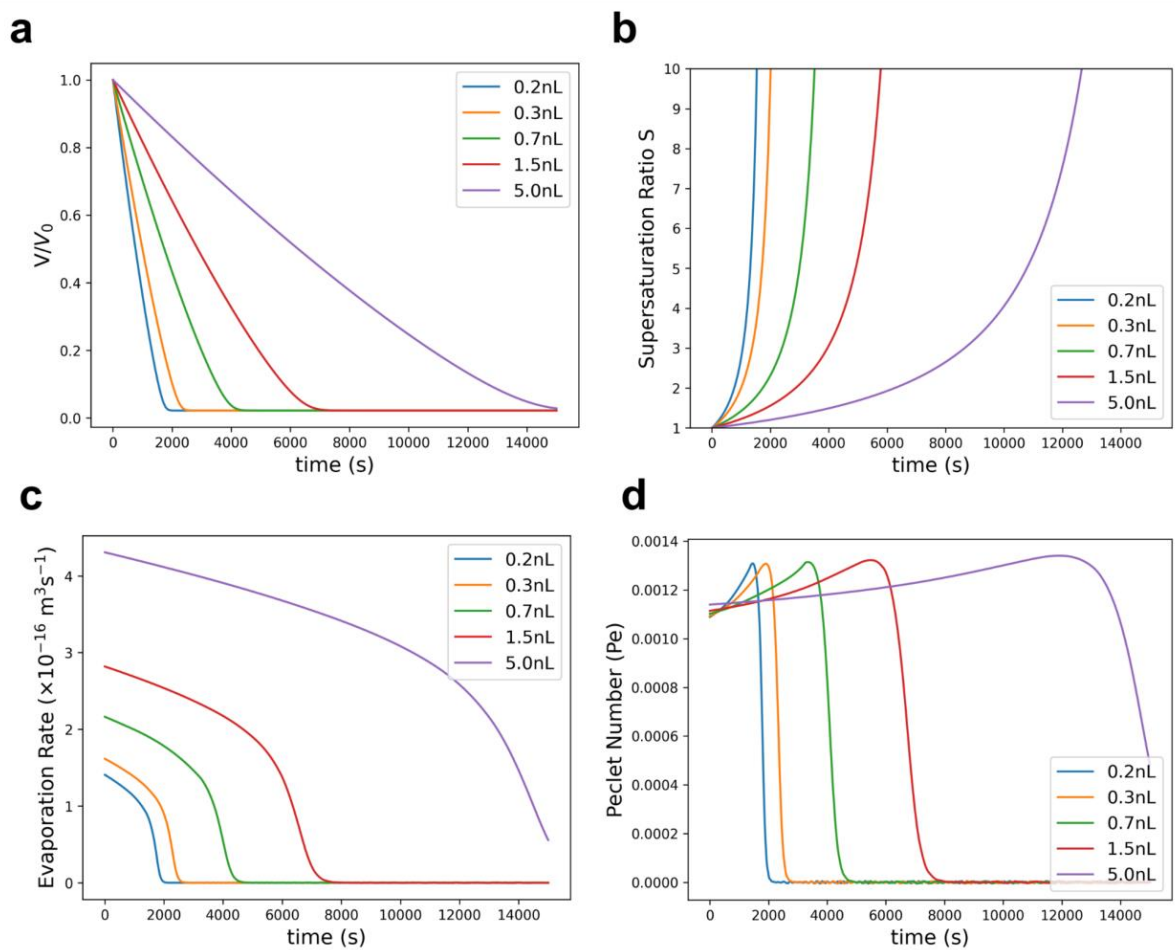

**Figure S2.** Model predictions for bottom-view arrays of microdroplets in terms of (a) relative volume $V/V_0$ (b) supersaturation ratio (c) evaporation rate (d) Peclet number[1, 8]. Pe < 1 suggests a uniform distribution of concentration within the droplet. The droplets were subjected under the following conditions: RH = 0.10, T = 25°C, P = 1 atm.